\documentclass[12pt]{article}
\pdfoutput=1
\usepackage{putex}
\usepackage{feyn}
\usepackage[vcentermath]{youngtab}
\usepackage{subfig}
\usepackage{lscape}
\usepackage{cite}
\usepackage{graphicx}
\usepackage{epstopdf}
\usepackage{enumerate}
\usepackage{tensor}
\usepackage{slashed}
\usepackage{amsmath}
\usepackage{amssymb}
\usepackage{mathrsfs}
\usepackage{lgrind}

\usepackage{amsmath,amssymb,braket}

\usepackage{bbm}

\usepackage{hyperref}

\numberwithin{equation}{section}

\newcommand {\be} {\begin {equation}}
\newcommand {\ee} {\end {equation}}

\newcommand {\bes} {\begin {equation*}}
\newcommand {\ees} {\end {equation*}}

\newcommand{\eps}{\epsilon}

\def\CH{{\cal H}}

\newcommand{\beq}{\begin{equation}}
\newcommand{\eeq}{\end{equation}}

\def\be{ \begin{equation} }
\def\ee{ \end{equation} }

\def \be {\beta}

\def \beq { \begin{equation}}
\def \eeq {\end{equation}}

\usepackage{indentfirst}
\begin{document}
\preprint{PUPT-2542}

\institution{PU}{Department of Physics, Princeton University, Princeton, NJ 08544}

\authors{Fedor K. Popov \worksat{\PU}}

\title{Debye mass in de Sitter space}

\abstract{
We calculate the one-loop contributions to the polarization operator for scalar quantum electrodynamics in different external electromagnetic and gravitational fields. In the case of gravity, de Sitter space and its different patches were considered. It is shown that the Debye mass appears only in the case of alpha-vacuum in the Expanding Poincare Patch. It can be shown either by direct computations or by using analytical and causal properties of the de Sitter space. Also, the case of constant electric field is considered and the Debye mass is calculated.}

\date{}
\maketitle
\tableofcontents
\section{Introduction}

One of the problems of the modern theoretical physics is the understanding of quantum field theory in the external gravitational or electromagnetic fields. These two problems at the first glance seem to be very different, but they share a lot of similarities and common properties. For instance, particle production occurs in both situations \cite{Schwinger:1951nm,Birrell:1982ix}.  Moreover, it is well-known that the constant electric field and de Sitter space-time share a lot of common infrared properties. Thus, both of them acquire large secularly growing contributions \cite{Krotov:2010ma,Polyakov:2012uc,Polyakov:1982ug,Polyakov:2007mm,Polyakov:2009nq,Akhmedov:2012pa,Akhmedov:2014doa,Akhmedov:2013xka,Akhmedov:2013vka,Akhmedov:2014doa,Akhmedov:2014hfa,Akhmedov:2017hbj,Anderson:2013zia,Anderson:2017hts,Woodard:2014jba}, that distort the tree-level picture by loop effects and can possibly destroy the initial background. 

In the case of gravity it is extremely difficult even to solve a free field theory for the general gravity background, but in the case of highly symmetric space-times like de Sitter calculations simplify and in principle can be done. Also, because the de Sitter space-time perhaps well describes the universe immediately after the Big Bang, quantum field theory in such a background is a very interesting topic for physics community. One of the effects that makes de Sitter space-time different from the Minkowski is the appearance of temperature that is proportional to the Hubble constant of de Sitter space-time $T_{dS} = \frac{H}{2\pi}$. For example, if we release a detector in the de Sitter space-time that has some energy levels, eventually we get an excited detector with a energy distribution like thermal with a temperature $T_{dS}$ \cite{Birrell:1982ix}. It gives rise to some questions concerning the physical origin of this temperature. For example, it is well-known that in plasma the temperature gives photon a mass, that is called the Debye mass. Does this effect happen in de Sitter space-time? If yes, is the photon mass determined by the temperature $T_{dS}$, does it depend on the choice of the vacuum state? Also, the value of the photon mass can indicate different divergences that are emergent in the de Sitter space. For example, if the square mass is negative, we have a direct sign of instability --- any initial perturbation will be amplified. Or the mass can become very large or even UV divergent, it means that any charge will be eventually screened.  This question can be projected to  the problem of understanding of the stability of de Sitter space-time itself. Indeed, if the photon aquires a mass, a graviton can also get it. In the case of thermal state in a flat space-time this mass is negative that leads to the Jeans' gravitational instability \cite{Jeans1,Gorbunov:2011zz,Gorbunov:2011zzc}.  The same scenario can happen also in the case of the impulse or eternal electric field. Due to the particle production, photon in such a medium can acquire a non-zero mass. If this mass is negative or very large, the initial electric field will be destroyed.

Below we will discuss how to define the Debye and magnetic masses for a curved gravitational background and perform calculations for de Sitter space-time for different types of invariant vacua.  We will get that there is no Debye mass for the case of Bunch-Davies vacuum, but there is a non-zero photon mass for any alpha-vacuum or for the de Sitter broken phase, while magnetic mass is equal zero for any chosen initial state. The analogous calculations will be done for the case of the constant electric field and the impulse, it will be shown that the Debye mass diverges linearly with the growth of time in the both situations. The last statement is similar to the linear growth of the electric current in the Schwinger effect \cite{Krotov:2010ma,Akhmedov:2014doa,Akhmedov:2014hfa}

\section{Polarization operator in curved space-time}
In this section we will define Debye-mass for scalar quantum electrodynamics on a some gravitational background, that we specify later.
\beq
S = \int d^4 x \sqrt{-g}\left[-\frac14 F_{\mu\nu}^2 + \left|\left(\partial_\mu + i e A_\mu\right) \phi\right|^2 - m^2 \left|\phi\right|^2\right], \quad g_{\mu\nu} = a^2(t) \eta_{\mu\nu}.
\eeq

To simplify the calculations we assume that the gravitational field doesn't depend on spatial coordinates and the metric is diagonal. Also, because the system is not at the equilibrium (for example the external gravitational field may depend on time) we will use Schwinger-Keldysh diagrammatic technique to calculate the polarization operator for photon. In this technique the number of fields is doubled: for every field we have a classical field, that can be thought as its value, and the quantum part, that can be considered as a fluctuation and gives the information about the spectrum of the system \cite{kamenev2011field,pitaevskii2012physical}.  Despite the fact that this technique is computationally difficult, it gives some advantages in comparison to the Feynman one, e.g. that we can write down the classical equations of motion deformed by quantum effects for the classical field \cite{Boyanovsky:1999jh}:
\beq
 \Box_x A_\mu(x,t) + \int\limits^t_{-\infty} dt' d^{3} y \sqrt{-g(t')} \,\Pi_{\mu\nu'}(x,t|y,t') g^{\nu'\nu}(t') A_\nu(y,t') = 0, \label{eom2}
\eeq
where $\Box_x$ is the corresponding wave operator for the vector field under some gauge conditions in the given space-time, for simplicity we consider Coloumb gauge $\partial^\alpha A_\alpha  =0$, that explicitly breaks the internal coordinate reparamtrization invariance. 

The polarization operator in the eq. \eqref{eom2} has the following form:
\beq
\Pi_{\mu\nu}(x,t|y,t') =i \, \theta(t-t') \braket{\Psi|\left[J_\mu(x,t), J_\nu(y,t')\right]|\Psi},\label{polarc}
\eeq
where $J_\mu(x,t)$ is a current operator for the given quantum field theory and $\Psi$ is an arbitrary state of the system. If the system is invariant under spatial translations we make a Fourier transformation of the polarization operator \eqref{polarc}
\beq
\Pi_{\mu\nu}(q,t,t') = \int d^3 x\, e^{-i q y}\, \Pi_{\mu\nu}(x,t|x+y,t')\label{polar}
\eeq

The equations of motions \eqref{eom2} can be thought also as the application of Kubo formula \cite{kubo1957statistical}. Indeed, the polarization operator \eqref{polar} has a form of a current susceptibility of the media to the presence of external electromagnetic field $A_\mu(x,t)$. And we try to take into account the influence of the induced currents on the external electromagnetic field. 

If we assume that the field $A_\mu(x,t)$ changes slowly in comparison with the scalar field fluctuations or with the polarization operator $\Pi_{\mu\nu}(x,t|y,t')$ then we can assume that $A_\nu(x,t) \approx A_\nu(y,t')$ to obtain
\beq
 \Box_x A_\mu(x,t) + A_\nu (x,t)\int\limits^t_{-\infty} dt' d^{d-1} y \sqrt{-g(t')} \,\Pi_{\mu\nu'}(x,t|y,t') g^{\nu'\nu}(t') = 0, \label{eom3}
\eeq

The last term in this equation has the form similar to the mass term, but because it has a tensor structure, this mass will depend on the polarization. To understand better its tensor structure we can make the Fourier transformation  of \eqref{polar} in spatial directions and constraint it with the use of $SO(3)$ symmetry and Ward identities. Then we get
\begin{gather}
\Pi_{00}(q,t,t') = C_E(q,t,t') = \ddot{M}_E(q,t,t'),\quad \Pi_{0\alpha} = \dot{M}_E(q,t,t') \frac{q_\alpha}{q^2}, \notag\\
\Pi_{\alpha\beta} = - C_M(q,t,t') \left(\delta_{\alpha\beta} - \frac{q_\alpha q_\beta}{q^2}\right) + M_E(q,t,t') \frac{q_\alpha q_\beta}{q^4}.
\end{gather}
Where $\alpha,\beta$ indexes run over the spatial directions, dots stands for the covariant differentiation with respect to the time $\dot{f}(t) = \partial_t \left( a^{-2}(t) f(t)\right)$. Substiting these expressions into the equations of motion \eqref{eom2} with the use of the Coloumb gauge we get
\begin{gather}
\Box_p A_0(q,t) + A_0(q,t) \int \limits^t_{-\infty} dt' \sqrt{-g} g^{00} C_E(q,t,t')  = 0,\quad {\rm and} \notag\\
\Box_p A_\alpha(q,t) +  A_\alpha(q,t)\int \limits^t_{-\infty} dt' \sqrt{-g} g^{00} C_M(q,t,t') = 0, \label{fineqforpoten}
\end{gather}
where $\Box_p$ is an image of spatial Fourier transformation of the wave operator $\Box_x$, that was introduced above.

So we can separate equations for $A_\alpha$ and $A_0$ fields. Therefore we can introduce two notions of mass like we did above
\begin{gather}
m^2_{Deb}(t) = \lim_{q\to 0} \int \limits^t_{-\infty} dt' \sqrt{-g} g^{00} C_E(q,t,t'),\quad m^2_{mag}(t) = \lim_{q\to 0} \int \limits^t_{-\infty} dt' \sqrt{-g} g^{00} C_M(q,t,t') \label{masses} 
\end{gather}
If there is a limit at $t\to \infty$ for these functions we introduce
\[
m^2_{Deb,mag} = \lim\limits_{t\to\infty} m^2_{Deb,mag}(t) 
\]
And the equations \eqref{fineqforpoten} will be rewritten as
\begin{gather}
\Box_q A_0(q,t) + m^2_{Deb} A_0(q,t) = 0,\quad \Box_q A_\alpha(q,t) + m^2_{mag} A_\alpha(q,t) = 0.
\end{gather} 
That indicates the appearance of the mass for the electromagnetic field (because the dispersion relation for the photon has a gap). The first one corresponds to the electric, or just a Debye mass, while the second one corresponds to the appearance of magnetic mass. If the magnetic mass is not equal to zero, it leads to the superconductivity \cite{landau2013statistical}. 
If the mass \eqref{masses} vanishes, we can expand it at small $q$ the next term will proportional to the momentum $q$ and corresponds to the electric, magnetic susceptibility and the conductivity of the state. If we get, e.g. that $m^2_{Deb}$ is negative or very big, it means that the system either in the state of instability or any initial electromagnetic field will be effectively screened.

\section{General properties of magnetic and Debye masses in the curved background}
\begin{figure}
	\centering
	\includegraphics[scale=0.8]{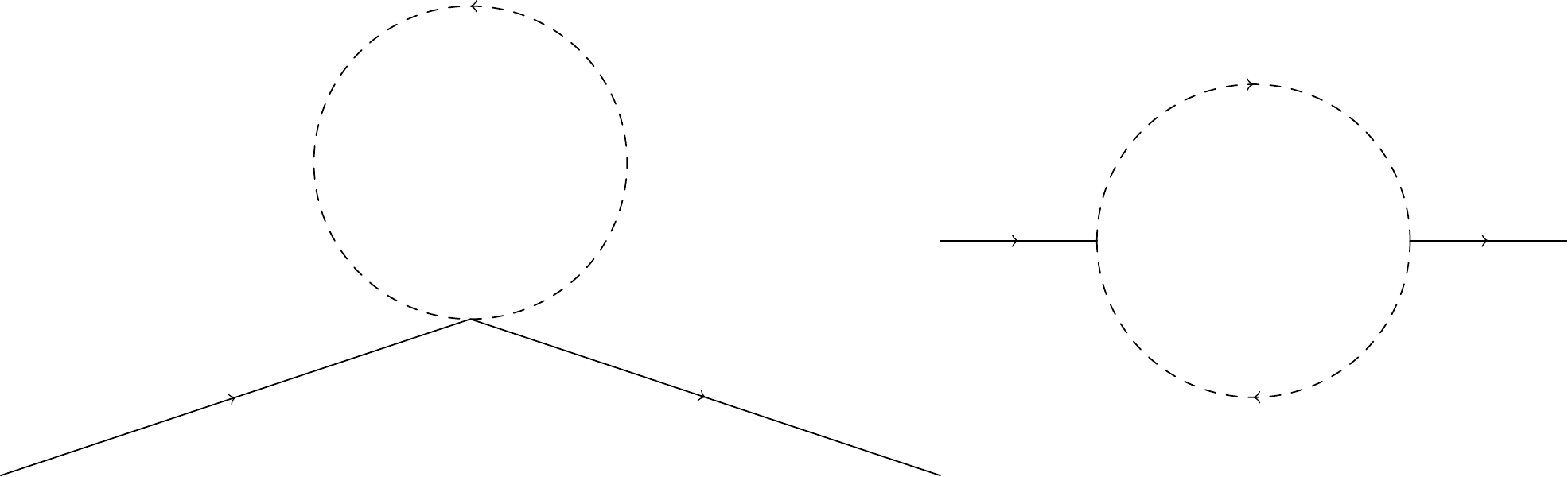}
	\caption{The one-loop contributions to the polarization operator. The dashed lines correspond to the scalar field. The solid ones are photon propagators.}
	\label{Diagrams}
\end{figure}
In this section we consider the general case of a gravitational background with $SO(3)$ spatial symmetry and discuss the general features of the magnetic and debye mass in the presence of an external gravitational field.
Let us choose some definition of the vacuum state and make the decomposition of the field $\phi$ through creation and annihilation operators
\beq
\phi = \int \frac{d^3 p}{(2\pi)^3} \left[a_p e^{i p x} f_p(t)  + b^\dagger_p f^*_p(t) e^{-i p x} \right],\quad [a_p, a^\dagger_q] = [b_p, b^\dagger_q] = (2\pi)^3 \delta(p-q).
\eeq
In order to satisfy the usual commutation relations the harmonics satisfy the Klein-Gordon equation
\beq
\left[\partial_t^2 + 2 \partial_t \log a(t) \partial_t + p^2 + m^2 a^{2}(t)\right]f_p(t) = 0. \label{eomharm}
\eeq
Further we will consider only the vacuum state, that is nullified by the annihilation operators $a_p\ket{\rm vac} = b_p \ket{\rm vac} = 0$. Then with the use of Keldysh-Schwinger diagrammatic technique we can calculate explicitly the magnetic and electric mass. To do it we need to sum up the contributions from two diagrams \cite{Boyanovsky:1999jh} (see the figure \ref{Diagrams}). One can show that the Debye mass doesn't get any UV divergences. Eventaully, we get the following expressions for masses according to the formulas \eqref{masses}
\beq
m^2_{\rm Deb} = \lim_{q\to 0}\, 2 e^2 \int^t_{-\infty} dt \sqrt{-g} g^{00} \int \frac{d^3 k}{(2\pi)^3} {\rm Im}\,\left[\left(f_{p+q}(t) \overleftrightarrow{\partial_t} f_{p}(t)\right) \left(f^*_{p+q}(t') \overleftrightarrow{\partial_t} f^*_{p}(t')\right)\right], \label{elemass}
\eeq
and for the case of magnetic mass we have
\beq
m^2_{\rm mag} \delta_{\alpha\beta} =8 e^2 \int \frac{d^{3} p}{(2\pi)^3} p_\alpha p_\beta \int\limits^t_{-\infty} dt' \sqrt{-g} g^{00} {\rm Im}\,\left[f^2_{p}(t) f^{*2}_{p}(t')\right] + 2 e^2 \delta_{\alpha\beta} \int \frac{d^{3} p}{(2\pi)^3} \left|f_p(t)\right|^2 \label{magmass}.
\eeq
Note that sometimes we can take the limit $t\to \infty$ to understand the behaviour of the system after the interaction, also this limit should be taken before taking the limit $q\to 0$, otherwise we get zero due to the charge conservation. For instance, in the case of quantum electrodynamics in the flat space in an excited state we have \cite{Migdal:1977bq}:
\begin{gather}
\Pi_{00}(q,\omega) \propto \int \frac{d^3 k}{(2\pi)^3} \frac{n_{k+q} - n_{k}}{\omega_{k+q} - \omega_{k} + \omega}, \quad \text{From where it follows that:}\notag\\
\quad \Pi_{00}(0,\omega) =0, \quad \lim_{q\to 0} \Pi_{00}(q,0)  \propto \int \frac{d^3 k}{(2\pi)^3} \frac{\partial n_k}{\partial \omega_{k}} \label{MigClass}
\end{gather} 
So formally, $\Pi_{00}(0,\omega) = 0$, but the Debye mass is not equal to zero, if $n_q$ is non-trivial. In general case we are not able to take the limit $t\to\infty$ because of the presence of the external fields. 

We can show that the magnetic mass is identical equal to zero. To do this we will use the equations of motions for the modes.
Then with the use of quantum mechanical perturbation theory one can prove the following identity
\beq
\partial_{p^2} \left|f_p(t)\right|^2 = 2\int\limits^t_{-\infty} dt' a^2(t') {\rm Im}\,\left[f^2_p(t) f_p^{*2}(t')\right], \label{qmiden}
\eeq
then  the equation \eqref{magmass} can be rewritten as 
\begin{gather}
m^2_{\rm mag} \delta_{\alpha\beta} 
= 2 e^2 \int \frac{d^{3} p}{(2\pi)^3} p_\alpha \partial_{p_\beta} p^2 \partial_{p^2} \left|f_p(t)\right|^2+ 2 e^2 \delta_{\alpha\beta} \int \frac{d^{3} p}{(2\pi)^3} \left|f_p(t)\right|^2 = \notag\\
= 2 e^2 \int \frac{d^3 p}{(2\pi)^3} p_\alpha \partial_{p_\beta} |f_p(t)|^2 + 2 e^2 \delta_{\alpha\beta} \int \frac{d^{3} p}{(2\pi)^3} \left|f_p(t)\right|^2 = \notag\\
=2 e^2 \delta_{\alpha\beta} \int \frac{d^3 p}{(2\pi)^3} |f_p(t)|^2 - 2 e^2 \delta_{\alpha\beta} \int \frac{d^{3} p}{(2\pi)^3} \left|f_p(t)\right|^2 = 0
\end{gather}
Where in the last line we have integrated by parts the first term and used the relation $\partial_{p_\alpha} p_\beta =\delta_{\alpha\beta}$. It would be interesting to consider the general case (no spatial $SO(3)$ and translational invariance, with off-diagonal components of metric) and investigate the conditions under which magnetic mass appears. Also it would be useful to prove that magnetic mass is equal to zero at all orders of perturbation theory. Due to the difficultity of this problem we will leave it for the future research.

Now let us consider the Debye mass. For the general case we can't prove that it is equal to zero. Nevertheless as it was shown above it can appear only due to the some singularities if we take the lower limit of the integration over $t'$ in \eqref{elemass} to $-\infty$. Indeed, let us rewrite the formula for the Debye mass as a commutator
\beq
m^2_{Deb} \propto  \lim\limits_{q\to 0} \int\limits^t_{-\infty} dt' \sqrt{-g} g^{00} \braket{\Psi| \left[J_0(t,q), J_0(t',-q)\right] |\Psi}
\eeq 
Because when we take the limit $q\to 0$ the operators under integrals become just total charges $\lim\limits_{q\to 0}J_0(t,q) \propto Q$, that trivially commutes with any operator due to the charge conservation we can easily shift the upper limit of the integration $t\to \tilde{t}$:
\beq
m^2_{Deb} \propto \lim\limits_{q\to 0} \int\limits^{\tilde{t}}_{-\infty} dt' \sqrt{-g} g^{00} \braket{\Psi| \left[J_0(t,q), J_0(t',-q)\right] |\Psi} \label{Debyepropery}
\eeq
So if the integral converges it means that there is no Debye mass in the system. For example, one can check that for Minkowski space-time and vacuum state harmonics behave like $f_p(t) \sim e^{- i \omega_p t}$ with some dispersion relation $\omega = \omega_p$. Such a behavior doesn't lead to singularities at $t \to - \infty$. While if there is a linear combination of exponents it can lead to the photon mass \cite{Boyanovsky:1999jh,Tsvelik:1996zj}. Indeed, consider
\begin{gather}
f_p(t) = \frac{1}{\sqrt{2\omega_p}}\left[\alpha_p e^{-i\omega_p t}+ \beta_p e^{i\omega_p t}\right],\quad \left|\alpha_p\right|^2 - \left|\beta_p\right|^2 = 1, \notag\\
m_{Deb}^2 = 2 e^2 \lim \limits_{q\to 0}\int \frac{d^3 p}{(2\pi)^3}  \frac{|\alpha_p|^2 |\beta_{p+q}|^2 - |\alpha_{p+q}|^2 |\beta_p|^2}{\omega_{p+q} - \omega_p} = - 2 e^2 \int\frac{d^3 p}{(2\pi)^3} \frac{\partial \left|\beta_p\right|^2}{\partial \omega_p}.
\end{gather}

It is known that $\left|\beta_p\right|^2$ can be interpreted as the number of the quasiparticles $n_p$ at the given level $p$. Then we obtain the classical formula for Debye mass (compare it to \eqref{MigClass})
\begin{gather}
m^2_{\rm Deb} =- 2 e^2 \int\frac{d^3 p}{(2\pi)^3} \frac{\partial n_p}{\partial \omega_p}. \label{DebMassMink}
\end{gather}
The minus sign shows that the photon mass is positive if $ \frac{\partial n_p}{\partial \omega_p}<0$, that can be considered as one of the necessary conditions for the system to be stable. If $\frac{\partial n_p}{\partial \omega_p} >0$ there can be some divergences and instabilities of the system.
\section{De-Sitter space-time}
In this section we consider different patches of de Sitter space-time such as Expanding Poincare Patch (EPP), Contracting Poincare Patch, Global de Sitter and one that was obtained from Minkowski space-time by switching on the eternal exponential expansion.
\subsection{Expanding Poincare Patch}
In the expanding patch of de Sitter space-time the metric has the following form \cite{Akhmedov:2013vka}
\beq
ds^2 = \frac{d\eta^2  - d\vec{x}^2}{\eta^2},\quad \eta\in\left(0,+\infty\right).
\eeq
Where we set the Hubble constant tp be one, $H=1$, and the conformal time changes from $+\infty$ to $0$ from the past to future infinity. Also we will only consider the case of massive fields in the de Sitter $m> \frac{D-1}{2}$, the case of light fields is similar, but a little bit harder to consider. The harmonics in such a case are linear combinations of Hankel's functions
\beq
f_p(\eta) = \eta^{\frac{d-1}{2}} h_{i\mu}(p\eta) = \eta^{\frac{d-1}{2}}\left[\alpha H^{(1)}_{i\mu}(p\eta) + \beta H^{(2)}_{i\mu}(p\eta)\right],\quad \left|\alpha\right|^2 - \left|\beta\right|^2 = 1. \label{HarmDsEPP}
\eeq
If $\beta=0$ then this vacuum is named after Bunch and Davies \cite{Bunch:1978yq}. This vacuum has some nice properties when the interactions are switched off. It can be thought as an analog of usual Minkowski vacuum in de Sitter space-time \cite{Krotov:2010ma,Akhmedov:2013vka}. It means that in the distant past ($\eta=\infty$) modes are just plane waves $e^{i p \eta}$. If $\beta\neq 0$ we will name such a state as an alpha vacuum \cite{Mottola:1984ar}. The last states are usually considered to be unphysical due to the incorrect UV behavior, nevertheless they are function of a geodesic distance $Z$ on a de-Sitter space and therefore could respect de Sitter invariance (at least on a tree level). Therefore, it is interesting to consider alpha-vacuums and compare to the Bunch-Davies(BD) state.

In terms of this modes the Debye mass is
\beq
m^2_{\rm Deb} = \lim \limits_{q\to 0} 2 e^2 \int\limits^\infty_\eta d\eta' \eta^3 \eta' \int \frac{d^3 p}{(2\pi)^3} {\rm Im}\left[\left(h_{i\mu}(p\eta)\overleftrightarrow{\partial_\eta} h_{i\mu}(|p+q|\eta)\right) \left(h^*_{i\mu}(p\eta')\overleftrightarrow{\partial_{\eta'}} h^*_{i\mu}(|p+q|\eta')\right)\right]. \label{massEPP}
\eeq
Before taking the limit $\eta\to 0$ one can notice that the LHS of the equation \eqref{massEPP} depends only on the physical momentum, $q\eta$, $m^2 (q\eta), m^2_{\rm Deb} = \lim\limits_{q\to 0} m^2(q\eta)$. Indeed, one can rescale $p\to p\eta, \eta'\to \frac{\eta'}{\eta}$ and get that the only left parameter is $q\eta$. In this case the limit of the distant future $\eta \to 0$ and small momentum $q\to0$  coincide with the limit of small physical momentum $q\eta \to 0$. As we will see, in the case of CPP there could arise some problems with defining of the Debye mass.

One can check that for Bunch-Davies vacuum there are no divergences as we integrate over $\eta'$, therefore there is no Debye mass $m_{\rm Deb,\, BD}^2  = 0$. In contrast, alpha-vacuums \eqref{HarmDsEPP} in the limit of $\eta\to\infty$ are combinations of the positive and negative frequency harmonics (partially, due to this fact they do not have a proper UV behavior) that leads to some divergences  
\beq
m^2_{\rm Deb} = -\frac{32 e^2}{\pi} \int \frac{d^3 p}{(2\pi)^3} {\rm Im}\left[\alpha^* \beta^* \left(h_{i\mu}(p) h'_{i\mu}(p) - p (h'_{i\mu}(p))^2 + p h_{i\mu}(p) h''_{i\mu}(p)\right)\right].
\eeq
It can be seen that the mass is divergent at the large $p$ and also may be negative for specific chosen alpha-vacuums. This indicates, that the alpha-vacuum can effectively screen any charge. The same effect can happen for the gravitational mass. In such a case, large mass can drastically change the de Sitter solution or even make it unstable.   

The difference between alpha vacuums and the Bunch Davies vacuum can be shown in the following way. In the BD vacuum propagators have a good analytic properties
\begin{gather*}
\hat{G}(X_1,X_2) = \begin{pmatrix}
G_{--} & G_{-+}\\
G_{+-} & G_{++}
\end{pmatrix} = \notag\\
=\begin{pmatrix}
G\left(\frac{\left(\eta_1 - \eta_2 - i \eps(\eta_1 - \eta_2)\right)^2 - \vec{x}^2}{\eta_1 \eta_2}\right) & G\left(\frac{\left(\eta_1 - \eta_2 - i \eps\right)^2 - \vec{x}^2}{\eta_1 \eta_2}\right)\\
G\left(\frac{\left(\eta_1 - \eta_2 + i \eps\right)^2 - \vec{x}^2}{\eta_1 \eta_2}\right) & G\left(\frac{\left(\eta_1 - \eta_2 + i \eps(\eta_1 - \eta_2)\right)^2 - \vec{x}^2}{\eta_1 \eta_2}\right)
\end{pmatrix}. \label{BDprop}
\end{gather*}
Note the $i\eps$ prescription in this equation. All two point functions for BD vacuums such as polarization operators will have the same $i\eps$ prescriptions for time, even after Fourier transformation over the spatial coordinates.
Now let us use this property in the formula for Debye mass in Kubo representation
\begin{gather}
m^2_{\rm Deb} =  \lim\limits_{k\to 0}\int\limits^\eta_{\infty} \frac{d\eta'}{\eta'^2} \braket{[J_0(k,\eta), J_0(-k,\eta')]}. \label{massphotoncoorEPP} 
\end{gather}
Because of the analytic properties discussed above this integral can be seen as the integral of the analytic function $\Pi_{00}(\eta') = \braket{J_0(k,\eta) J_0(-k,\eta')}$  of the variable $\eta'$, where the contour goes from $+\infty-i\eps$ to $\eta-i\eps$ turns up and goes from $\eta+i\eps$ to $+\infty+i\eps$. This allows us to make the integral \eqref{massphotoncoorEPP} convergent  by slightly changing the contour $\eta\to (1+i\delta) \eta$. Therefore we can interchange the integral and the limit to get 
\begin{gather}
m^2_{\rm Deb, BD} = \int\limits^\eta_{\infty} \frac{d\eta'}{\eta'^2} \braket{[J_0(0,\eta), J_0(0,\eta')]} = 0.
\end{gather}
This proof will not work for other vacuums, because the functions will not be analytic on the whole Riemann surface, for example, $\eta=\infty$ will have an essential singularity. Namely, the product of two Green functions for an alpha-vacuum contains two cuts and contour lies between them.
\begin{figure}
	\centering
	\includegraphics[scale=1]{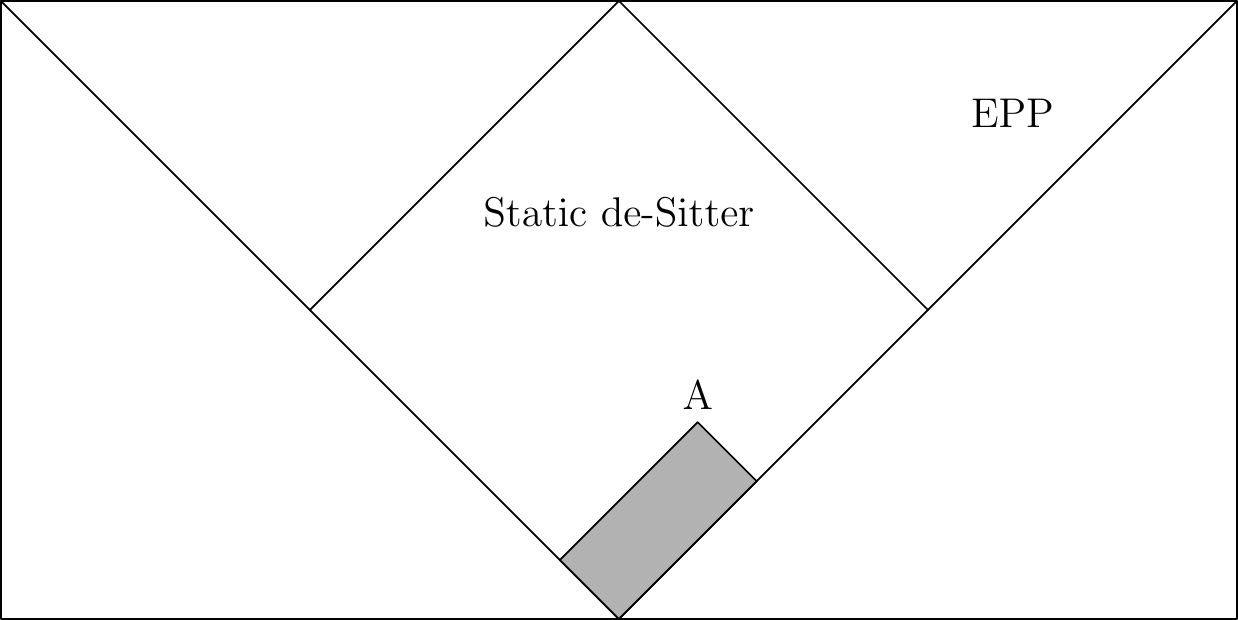}
	\caption{The contributions to the one-point correlation function at the point A come from the gray region. This gray region lays both within static de Sitter and EPP. Hence, for a proper choice of the state in static de Sitter the calculations should give the same answers.}
	\label{Static}
\end{figure}
Also the fact that the Debye mass for electric field is equal to zero for Bunch-Davies vacuum can be shown in the following way. The Bunch-Davies vacuum is the only state that respects the de Sitter isometry at any level of perturbation theory, while any alpha vacuum does break the global de Sitter symmetry even in the first loop \cite{Krotov:2010ma,Akhmedov:2013vka}. The proof will work for our situation, because of this if we get nonzero photon mass $m^2_{\rm Deb}$ to respect de Sitter isometry group and gauge invariance the equation for the classical electromagnetic field \eqref{eom3} should have the following form
\begin{gather}
\Box A_\mu  + m_{\rm Deb}^2\left(\delta^\nu_\mu - \nabla_\mu\cdot \Delta^{-1} \cdot \nabla^\nu\right) A_\nu=0
\end{gather}  
Therefore we should get the non-zero magnetic mass, but as it was shown above it is equal to zero for any choice of gravitational background. 
The similar calculation was done in a locally de Sitter space-time for a nearly minimal coupled, light scalar field \cite{Prokopec:2003tm,PhysRevLett.89.101301,Prokopec:2003bx}, but the conclusions are different.

\subsection{Static coordinates}
In this section we consider the static patch of EPP and calculate the Debye mass for the vacuum state (see figure \ref{Static}). Because the Keldysh-Schwinger diagrammatic technique is causal, the calculation in static de Sitter and EPP should coincide for a proper choice of the state. We do not think, that the BD state in the EPP corresponds to the vacuum state in the static coordinates. In some sense the static de Sitter is similar to Rindler space. In the latter case the thermal propagator in Rindler space corresponds to the usual Minkowski propagator \cite{Moschella:2009ub}, therefore one would except the emergence of photon mass in Rindler space, while in the Minkowski space photon mass of the vacuum state must be equal to zero. Therefore it is interesting to investigate the difference between this situations also in the case of the de Sitter space. Because the thermal propagator for the static de-Sitter is unknown, we consider only the case of the vacuum state of static de Sitter.

The metric of static de Sitter is
\beq
ds^2 = \left(1 - r^2\right) dt^2 -\frac{1}{1 - r^2} dr^2 - r^2 \left(d\theta^2 + \sin^2 \theta d\phi^2\right) \label{static}
\eeq
The equations of motion for modes are
\begin{gather}
\left(\frac{1}{1-r^2} \partial_t^2 - \frac{1}{r^2}\partial_r \left(r^2(1-r^2) \partial_r\right) - \frac{\Delta_{\theta,\phi}}{r^2} \right)f_{l,m,\omega}(t,r,\theta,\phi) = 0
\end{gather}
The modes behave as
\beq
f_{l,m,\omega} = \frac{1}{\sqrt{2\omega}} e^{-i \omega t} Y_{l,m}(\theta,\phi) \label{staticharm} h_{\omega,l,m}(r),
\eeq
where $h_{\omega,l,m}(r)$ satisfies the following equation
\beq
\left(-\frac{1}{1-r^2} \omega^2 - \frac{1}{r^2}\partial_r \left(r^2(1-r^2) \partial_r\right) + \frac{l(l+1)}{r^2} \right)h_{\omega,l,m}(r) =0,
\eeq
these modes are complex conjugated to each other $h^*_{\omega,l,m} = h_{-\omega,l,m}$ and are orthogonal
\beq
\int\limits^1_0  \frac{r^2 dr}{1-r^2} h_{\omega,l,m}(r)h^*_{\omega',l,m}(r) \propto \delta(\omega-\omega') \label{ortho}.
\eeq
The modes in the eq. \eqref{staticharm} corresponds to the Bunch-Davies vacuum, if instead we consider a linear combinations we get modes for alpha-vacuums. Then we can get the formula for the polarization operator in a mixed representation for $L,M=0,\Omega=0$ 
\begin{gather}
\Pi^R_{00}(\Omega=0,r,r') = 2 e^2 \sum_{l,m} \int \frac{d\omega}{2\pi} \,{\rm Im}\,\left[h^2_{\omega,l,m}(r) h^{*2}_{\omega,l,m}(r')\right]
\end{gather}
To get the actual photon mass we need to integrate the above expression over the space with measure $\sqrt{-g} g^{00} \sim \frac{r^2}{1-r^2}$ (compare to the derivation of the eq. \eqref{elemass})
\beq
m^2_{\rm Deb} \propto \int \frac{ r^2 dr}{1 - r^2}\Pi^R_{00}\left(\omega=0,r,r'\right) = 2 e^2  \int \frac{r^2 dr}{1 - r^2} \sum_{l,m} \int \frac{d\omega}{2\pi} \,{\rm Im}\,\left[h^2_{\omega,l,m}(r) h^{*2}_{\omega,l,m}(r')\right]=0\notag
\eeq
Where the last equation comes from the orthogonality conditions \eqref{ortho}. While this calculation shows that photon mass in the vacuum state of static coordinates is equal to zero, BD vacuum may not correspond to the vaccuum state. 

\subsection{Contracting Poincare Patch}
For the case of the Contracting Poincare Patch we have the same modes and the same expression for Debye mass of photon \eqref{massEPP}, but the integration over time goes from 0 (the time is reserved in the case of CPP). So, we have
\beq
m^2_{\rm Deb} = \lim \limits_{q\to 0} e^2 \int\limits^\eta_0 d\eta' \eta^3 \eta' \int \frac{d^3 p}{(2\pi)^3} {\rm Im}\left[\left(h_{i\mu}(p\eta)\overleftrightarrow{\partial_\eta} h_{i\mu}(|p+q|\eta)\right) \left(h^*_{i\mu}(p\eta')\overleftrightarrow{\partial_{\eta'}} h^*_{i\mu}(|p+q|\eta')\right)\right].
\eeq
One can check that there are no additional divergences in the lower limit of integration over $\eta'\to 0$, because all harmonics behave well as the argument goes to zero. As it was discussed before eq. \eqref{Debyepropery} in such a case we can set $\eta = 0$ and get immediately that $m^2_{\rm Deb} = 0$.

Also there is a problem slightly mentioned in the previous subsection. Again before taking the limit $q\to 0$ we see that the mass is a function of the product of momentum and time $m^2_{\rm Deb}= m^2(q\eta)$. And the limits of the distant future and the small momentum are different. In order to prevent such an ambiguity we take the limit of small $q$ and don't take the limit of big $\eta$. Immediately, we get that this limit is equivalent to the case when we keep $q$ fixed and send $\eta\to 0$. That gives us that $m^2_{\rm Deb} = 0$. The limit when we take $q\eta$ to be equal infinity, corresponds to the EPP and will give us the same answer as in the previous subsection.

In the previous section we showed that there is no photon mass for Bunch-Davies vacuum, while for an alpha-vacuum there is a photon mass. One can ask what is a crucial difference between these two cases? The answer is the following. In the case of CPP, the causal structure is quite simple. There is only a compact set of points, that are causally connected to a given point. It leads to the absence of photon mass. Indeed, the coordinate representation of the Debye mass is as usual
\beq
m^2_{\rm Deb} \propto \int d^3 x \int\limits^\eta_0 \frac{d\eta'}{\eta'^2} \braket{[J_0(0,\eta), J_0(\vec{x},\eta')]}.
\eeq
Also there is no singularity on the lower limit of the integration over $\eta'$, because in this case the correlator goes to the zero. Now we have to integrate over spatial coordinates only in a ball of radius $\eta$, because only there lie points causally connected to the point of observation $\left(\eta,\vec{x}\right)$. Therefore we get
\begin{gather}
m^2_{\rm Deb} \propto \int \limits_{|\vec{x}| < \eta} d^3 x \int\limits^\eta_0 \frac{d\eta'}{\eta'^2} \braket{[J_0(0,\eta), J_0(\vec{x},\eta')]} =\int\limits^\eta_0 \frac{d\eta'}{\eta'^2}  \int \limits_{|\vec{x}| < \eta'} d^3 x \braket{[J_0(0,\eta), J_0(\vec{x},\eta')]} =\notag\\
=\int\limits^\eta_0 \frac{d\eta'}{\eta'^2}  \int d^3 x \braket{[J_0(0,\eta), J_0(\vec{x},\eta')]} = \int\limits^\eta_0 \frac{d\eta'}{\eta'^2}  \int d^3 x \braket{[J_0(0,\eta),Q]} =0 \label{CPPans}
\end{gather}
The integrand of \eqref{CPPans}  is a continuous function without singularities, because we integrate over a compact space we can interchange integrals over time and space, and get zero due to the charge conservation. However, if we calculate the higher loops, there are IR divergences \cite{Krotov:2010ma,Akhmedov:2013vka}. They may break the above proof. Indeed, we can simply interchange the integrals over space and time, because we do not have divergences on the lower limit of the integration over $\eta'$.

The conclusions about different photon masses in the case of EPP and CPP cases may be surprising, because as it was said above these patches are different only by the direction of time arrow. Nevertheless, this small difference brings a huge discrepancy between the dynamics of the systems. E.g., in the CPP the initial Cauchy surface is space-like, while in the EPP it is light-like, in the CPP there are IR divergences, while in the EPP there is only large IR contributions \cite{Krotov:2010ma,Akhmedov:2013vka}. 

Also, one can state that any point in the EPP lays in some CPP, so we should conclude that $m^{EPP}_{Deb}=m^{CPP}_{Deb} = 0$ even for alpha-vaccum. This contradiction is easy to resolve. When we get the $m^{EPP}_{Deb}$ we indeed integrate the same propagators, but not over the whole space, but only over the half of the space. So it can lead to the different results.
\subsection{Global De-Sitter}
In the case of Global De-Sitter we have a problem with defining a photon mass. This can be seen as follows. The spatial cross-section is a three-dimensional sphere $S^3$, that is compact. Hence, the spatial momentum of an electric potential $A_0(x,t)$ is discrete and numerates by $\vec{M}$ with the eigenfunctions $Y_{\vec{M}}\left(\vec{x}\right)$. Because there is a gauge transformation $A_0(x,t) \to A_0(x,t) + f(t)$, where $f(t)$ is arbitrary, it is easy to see that the case of $\vec{M} = \vec{0},Y_{\vec{0}}\left(\vec{x}\right)=1$ corresponds to the gauge degrees of freedom. It gives that physical photons must have a non-zero spatial momentum and we can not see whether there is a gap or not. In such a case formally we have to assign a zero Debye mass in order to keep gauge invariance. In the case of Minkowski space-time we had a similar problem, that the photons with $\vec{k}=0$ are really unphysical, but we can investigate the spectrum near $\vec{k}$ because $\vec{k}$ is non-discrete. The above reasoning also gives us that the polarization operator should be equal to zero as we choose $\vec{M}=0$. More formally, the polarization operator is
\begin{gather}
\Pi^0_0(M,M',t_1,t_2) \propto \braket{\Psi|[Q_{M}(t_1),Q_{M'}(t_2)]|\Psi},
\end{gather}
where $Q_{M}$ is defined as
\[
Q_M(t) = \int \sqrt{\gamma}\, d^{d-1} x\,  Y_M(x)\, Q(x,t), 
\]
and $\gamma$ is a determinant of the spatial metric.

Because of $M$ is discrete, we cannot take the limit $M\to 0$ and that leads to
\begin{gather}
\Pi^0_0(0,0,t_1,t_2) \propto \braket{\Psi|[Q_{0}(t_1),Q_{0}(t_2)]|\Psi} = 0,
\end{gather}
because of the charge conservation.
Let us express the $\Pi^0_0(t,x|t',x')$ as a series
\begin{gather}
\Pi_{00}(t,x|t',x') = \sum_{\vec{M},\vec{M'}} Y_{\vec{M}}(x) Y_{\vec{M'}}(x') \Pi^0_0(\vec{M},\vec{M'},t,t')\label{PolCPP}
\end{gather}
Where $\vec{M}$ and $\vec{M'}$ numerate modes on the spatial section. The case $\vec{M}=0$ corresponds to the zero mode, that is just a constant. Again, it is easy to see that such a term in the eq. \eqref{PolCPP} vanishes. In fact,
\begin{gather}
\sum_{\vec{M'}} Y_{\vec{M'}}(x') \Pi^0_0(0,\vec{M'},t,t') = \int \sqrt{\gamma}d^{d-1} x \Pi^0_0(t,x|t',x') Y_0(x) \propto \int \sqrt{\gamma} d^{d-1} x\braket{\Psi|[J_0(x,t), J_0(x',t')] |\Psi} = \notag\\
=\braket{\Psi|[Q, J_0(x',t')] |\Psi} = 0.
\end{gather}
Where we have used that the total charge of the system is equal to zero $Q\ket{\Psi} = 0$. It can happen if and only if $\Pi^0_0(0,\vec{M'},t,t') \equiv 0$. Therefore we get that in the eq. \eqref{PolCPP} $\vec{M} \neq 0$ and $\vec{M}' \neq 0$. As we discussed above there is no such a notion as a Debye mass in the compact case, but we can formally check that the gauge degrees of freedom does not acquire a mass. So we can follow the usual procedure and integrate over $t'$ to calculate the mass
\beq
m^2_{\rm Deb} = \int d^{d-1} x \sum_{\vec{M}\neq 0,\vec{M'} \neq 0} Y_{\vec{M}}(x) Y_{\vec{M'}}(x') \int^t_{-\infty} dt' a^{d-1}(t')\Pi^0_0(\vec{M},\vec{M'},t,t') = 0,
\eeq
where we have used that $\int d^{d-1} x\, Y_{\vec{M}}(x) = 0$ as $\vec{M} \neq 0$. The difference with Minkowski space-time is that the index in this case is discrete, which allows us to interchange the integrals over time and space.

\begin{figure}
	\centering
	\includegraphics[scale=0.8]{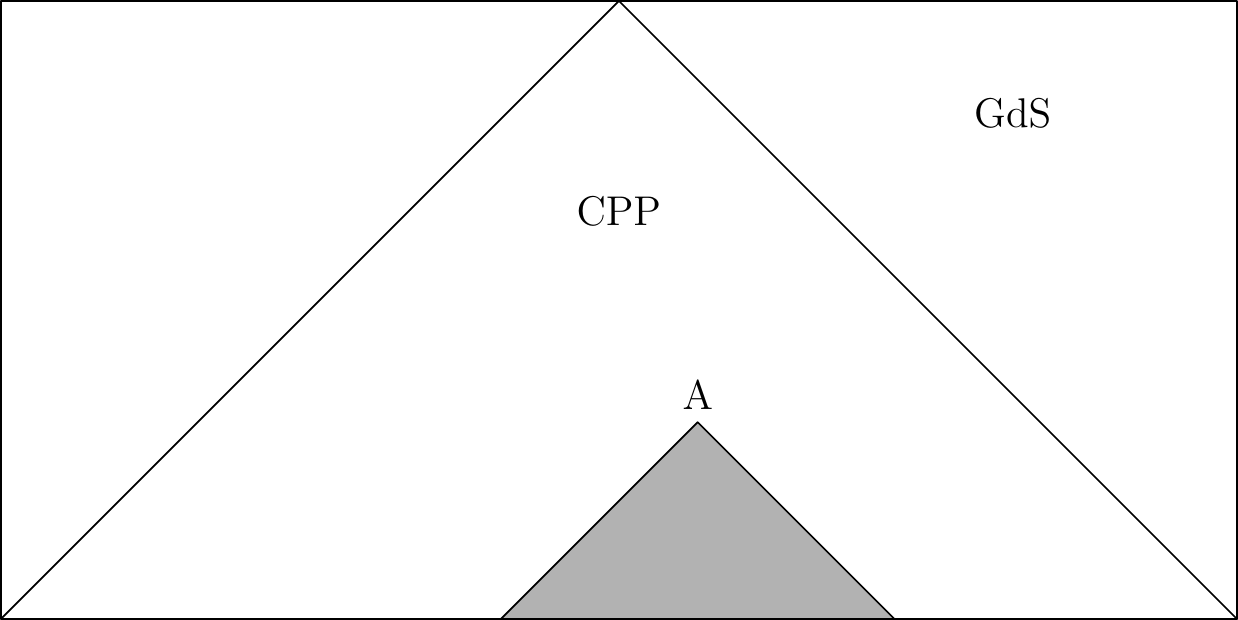}
	\caption{The Penrose Diagram for the two-dimensional de Sitter space-time. The shaded region corresponds to the causal connected points with the given point $A$. It is easy to see that the whole shaded region  lies in a Contracted Poincare Patch.}
	\label{Arguemnt}
\end{figure}

The above statement that may indicate that the calculations in the Contracting Poincare Patch and in the Global De-Sitter are in the agreement with each other \cite{Krotov:2010ma}. Indeed, one can prove that if one uses the Keldysh-Schwinger diagrammatic technique and calculates the correction to an one-point function, then there is no difference between these two calculations. The proof is quite simple,the  Keldysh-Schwinger technique is causal in contrast with the Feynman one, therefore all contributions only can come from the past light cone, but for a point in Global-de Sitter its causal past lays within a CPP (see the figure \ref{Arguemnt}). It proves that the calculation of one-point functions in the CPP and in Global de Sitter must be in agreement with each other.

\subsection{Breaking De-Sitter invariance}
In this subsection we consider the case when the de Sitter symmetry is broken. Namely we consider a situation with an inflation described by the Expanding Poincare Patch, but the scale parameter $a(t)$ is switched off in a while
\begin{gather}
a(t) = e^{T \tanh(\frac{t}{T}) - T}
\end{gather}
After the expansion phase the system will be in an excited state. We can find this state by gluing modes at the moments when de Sitter was switched off and on (see the paper \cite{Akhmedov:2017dih}). We get the following level population in the limit $T\to \infty$
\begin{gather}
n_p = \left\{\begin{matrix} 0, |p| > \mu, |p|< \mu e^{-2 T}\\
\frac{1}{e^{2\pi \mu} - 1}, \mu e^{-2 T} < |p| < \mu
\end{matrix}
\right.
\end{gather}
where $\mu = \sqrt{m^2 - \left(\frac{D-1}{2}\right)^2}$ and $m$ is a mass of the scalar field.

In such a case we can use the above formula for Minkowski space-time that gives us
\begin{gather}
m^2_{\rm Deb} = \frac{e^2 \mu^2}{2\pi^2} \frac{1}{e^{2\pi \mu} - 1}
\end{gather}
The formula is different from the Debye mass for a thermal state with Gibbons-Hawking temperature $T = \frac{1}{2\pi}$, it happens because of the infinite blue-shift during the expansion phase.

One can expect then, that we should get the same answer for the case of the CPP  with the contraction switched off after a while. However this means that we will get a contradiction with the above calculation of the Debye mass in CPP, because the same reasoning as in that case leads to the zero photon mass. This contradiction is easy to resolve. Indeed, we know that the limits of small $q$ and distant future do not commute. But in the case when we switch off the contraction there is no such a problem and we can take the limit of the distant future and get the non-zero photon mass. 

Let us consider the case when the EPP was turned on at some moment $\eta_0$. In this case after gluing modes will be described as follows
\beq
f_p(\eta) = \left\{
\begin{matrix}
H^{(1)}_{i\mu}(p\eta),\quad p\eta_0 \gtrsim \mu\\
J_{i\mu}(p\eta), \quad p\eta_0 \lesssim \mu
\end{matrix}
\right.
\eeq
As one can notice, there is no singularity from the switching on the EPP, effectively if one changes all momentum to the physical ones in eq. \eqref{massEPP}, he gets that the border between two states goes down $\mu' = \mu \frac{\eta}{\eta_0} \to 0, \eta \to 0$. So, we would expect that the state will be described by Bunch-Davies and that will not give any photon mass.  
\section{Debye mass in strong electric field}
In this section we derive the photon mass for Schwinger process. We assume that the electric field was never turned on, i.e. it always exists. The situation is similar to the de Sitter one, where we assumed the eternal inflation. And indeed, the IR properties in this two situation are similar and were discussed in many papers \cite{Krotov:2010ma,Polyakov:2012uc,Akhmedov:2014doa,Akhmedov:2014hfa}.

We choose the spatial gauge $A_1 = - E t, A_\perp = 0$. In this case the modes obeying the following equations
\begin{gather}
\left[\partial_t^2 + m^2 +p_\perp^2 + \left(p_1 + e E t\right)^2 \right] f_p(t) = 0, \quad f_p(t) = f_{p_\perp}(p_1+ e E t).
\end{gather} 
Solutions to this are parabolic cylindric functions. In this case we cannot define modes that are similar to the Bunch-Davies ones in EPP, because in the limit of large times modes are not simple plane waves. Therefore we choose the following boundary conditions \cite{Krotov:2010ma,Akhmedov:2014hfa}:
\begin{gather}
f_{p_\perp}(p_{\rm phys}) = \left\{\begin{matrix}
\frac{\alpha}{\sqrt{2p_{\rm phys}}} e^{- i \frac{p^2_{\rm phys}}{2 e E}}+\frac{\beta}{\sqrt{2p_{\rm phys}}} e^{ i \frac{p^2_{\rm phys}}{2 e E}}, \quad p_{\rm phys} \lesssim 0,\\
\frac{\gamma}{\sqrt{2p_{\rm phys}}} e^{- i \frac{p^2_{\rm phys}}{2 e E}} + \frac{\delta}{\sqrt{2p_{\rm phys}}} e^{i \frac{p^2_{\rm phys}}{2 e E}}, \quad p_{\rm phys} \gtrsim 0.
\end{matrix}\right. \notag\\
\text{where}\quad \left|\alpha\right|^2 - \left|\beta\right|^2 = \left|\gamma\right|^2 - \left|\delta\right|^2 = 1 \label{electricharmonics}
\end{gather}
The photon mass in such a case can be calculated in the similar way and the result is:
\begin{gather}
m^2_{\rm Deb} = - 2 e^2 \lim_{q\to 0}\int\limits^t_{-\infty} dt' \int \frac{d^3 p}{(2\pi)^3} {\rm Im}\left[\left(f_{p_\perp +q_\perp}(p_1+q_1 + e E t) \overleftrightarrow{\partial_t} f_{p_\perp}(p_1 + e E t)\right)\right.\times\hspace{5cm}\label{elephotmass}\\\hspace{7cm}\times\left. \left(f^*_{p_\perp +q_\perp}(p_1+q_1 + e E t') \overleftrightarrow{\partial_{t'}} f^*_{p_\perp}(p_1 + e E t')\right) \right]. \notag
\end{gather}
It is easy to see that there are no divergences as $q_\perp \to 0$, the possible divergence may appear only when $q_1 \to 0$. One can notice that the dependence on time $t$ can be removed. Indeed let redefine in \eqref{elephotmass} all momentums into the physical ones $p_{\rm phys} = p_1 + e E t$. Then we have 
\begin{gather}
m^2_{\rm Deb} = - 2 e^3 E  \lim_{q_1\to 0}\int\limits^0_{-\infty} dt' \int \frac{d^3 p}{(2\pi)^3} {\rm Im}\left[\left(f_{p_\perp}(p_1+q_1) \overleftrightarrow{\partial_{p_1}} f_{p_\perp}(p_1)\right)\right.\times\hspace{5cm}\notag\\\hspace{7cm}\times\left. \left(f^*_{p_\perp}(p_1+q_1 + e E t') \overleftrightarrow{\partial_{t'}} f^*_{p_\perp}(p_1 + e E t')\right) \right].
\end{gather} 
Now we can calculate the photon mass. The integral has a singularity
\begin{gather}
\int\limits^0_{-\infty} dt'  f^*_{p_\perp}(p_1+q_1 + e E t') \overleftrightarrow{\partial_{t'}} f^*_{p_\perp}(p_1 + e E t') = \frac{2\alpha^*\beta^*}{q_1} + \mathcal{O}(q_1), q_1 \to 0.
\end{gather}
That eventually gives us
\beq
m^2_{\rm Deb} = -4 e^2 (eE)  \int \frac{d^3 p}{(2\pi)^3} {\rm Im}\left[\alpha^* \beta^* \left(f'_{p_\perp}(p)^2 -f_{p_\perp}(p)f''_{p_\perp}(p)\right)\right].
\eeq
Where a prime corresponds to the differentiation with respect to the $p_1$. One can notice that this integral actually diverges quadrically, when $p_1 \to \infty$. It shows again that the eternal electric field is unphysical. Namely, a perturbation of the electric field may lead to the decay to another state \cite{Akhmedov:2014doa,Akhmedov:2014hfa}. 

To understand the meaning of this divergence we consider the case when the electric field was turned on at some moment. So we have the electric impulse: the field was switched on, worked for a time $T$ and turned off after. The only modes that have crossed the horizon are excited \cite{Polyakov:2012uc}. It will give the following level population \cite{Schwinger:1951nm}:
\beq
n_p = \left\{\begin{matrix}
	0, p_1 \gtrsim 0, \quad p_1 \lesssim -e E T\\
	e^{-\frac{\pi}{eE}(m^2+p_\perp^2)}, -e E T \lesssim p_1 \lesssim 0 ,
\end{matrix}\right.
\eeq
and will allow us to use the eq. \eqref{DebMassMink}.
This formula comes from the gluing of modes when the field was turned on and switched off. Then the mass is
\begin{gather}
m^2_{\rm Deb} = -2 e^2 \int \frac{d^3 p}{(2\pi)^3} \frac{\partial n_p}{\partial p_i} \frac{\partial p_i}{\partial \omega_p} = - 2 e^2 \int \frac{d^3 p}{(2\pi)^3} \frac{\partial n_p}{\partial p_i} \frac{\omega}{p_i} \approx - 4 e^2 \int\limits^0_{-e E T} \frac{dp_1}{2\pi} \int \frac{d^2 p}{(2\pi)^2} \omega \frac{\partial n_p}{\partial p_\perp^2} \approx \notag\\
\approx - 4 e^2 \int\limits^0_{-e E T} \frac{dp_1}{2\pi} p_1 \int \frac{d p_\perp^2}{4\pi}\frac{\partial n_p}{\partial p_\perp^2} = \frac{e^2}{4 \pi^2} e^{-\frac{\pi m^2}{e E}} (e E T)^2.
\end{gather} 
Where we extracted the leading contribution in the limit $T\to \infty$. As one can see the Debye mass indeed diverges quadratically with the growth of the length of the impulse.
Eventually it gives the following expression for photon mass after the impulse
\begin{gather}
m_{\rm Deb} = \frac{e^2}{2\pi} e^{-\frac{\pi m^2}{2 e E}} \left|A_1(\infty) - A_1(-\infty)\right|.
\end{gather}
This mass may be large for sufficiently long impulse. It means that any electric field is screened, including the initial impulse. That mean a long electric impulse is a unphysical situations and in order to understand the better the situation we have to take into account the backreaction of the scalar field on the electromagnetic field.
\section{Conclusion}
In this paper we discuss  the Debye and magnetic photon masses in the framework of particle creation by the external electromagnetic and gravitational fields. We argue that for the alpha vacuums in de Sitter space-time there is a non-zero Debye mass, while it is vanishing for the Bunch-Davies vacuum, which is an analog of the usual Minkowski vacuum. The last fact can be considered as consequence of the analytical properties of Green functions for Bunch-Davies vacuum. These properties also lead to the de Sitter invariance of the loop corrections in the Expanding Poincare Patch \cite{Krotov:2010ma}. 

Also, the cases of Global De-Sitter and Contracting Poincare Patch were considered. It was shown that for any chosen vacuum there is no photon mass.  These observation lead to the question of the physical meaning and relevance of the Gibbons-Hawking temperature. The same calculations were done for the eternal electric   
field and impulse, that features the appearance of a very large Debye mass, that can indicate that the considered situations are unphysical. In this paper we did not calculate the higher corrections to the photon mass and used only initial propagators. As it was shown \cite{Krotov:2010ma,Akhmedov:2013vka} these propagators receive large IR contributions. It would be interesting to investigate the effect of these large IR loop contributions to the Debye mass. Also it would be interesting to consider gravitational mass and its impact on quantum field systems. 
\section*{Acknowledgments}
The work is devoted to the memory of V.N.Diesperov.

I am grateful to  E.T. Akhmedov, I.R. Klebanov and A.M.Polyakov  for important discussions and remarks.  The work was supported in part by the US NSF under Grant No. PHY-1620059. 

\bibliography{mybib}{}
\bibliographystyle{ssg}

\end{document}